\documentstyle[prl,epsbox,epsf,aps]{revtex}
\begin{document} 
\draft
\title{Semiclassical theory of $h/e$ Aharonov-Bohm oscillation \\for doubly connected 
ballistic cavities}
\author{Shiro Kawabata}
\address{
Physical Science Division, Electrotechnical Laboratory, Umezono 1-1-4, Tsukuba, 
Ibaraki 305, Japan \thanks {E-mail:shiro@etl.go.jp}
}
%
\maketitle
\begin{abstract} 
In Aharonov-Bohm (AB) cavities forming doubly connected ballistic structures, $h/e$ AB oscillations that result from the 
interference among the complicated trapped paths in the cavity can be described by
the framework of the semiclassical theory.
We derive formulas of the correlation function $C(\Delta \phi)$ of the nonaveraged magnetoconductance for chaotic and regular AB cavities.
The different higher harmonics behaviors for $C(\Delta \phi)$ are related to the differing distribution of classical dwelling times.
The AB oscillation in ballistic regimes provides an experimental probe of quantum signatures of classical chaotic and regular dynamics.

\end{abstract}
%
%
%
Electron transport through quantum cavities is an exceedingly rich experimental system, bearing the quantum signature 
of chaos. \cite{rf:QC}
On the theoretical form, powerful techniques based on semiclassical approaches have produced specific 
predictions testable by experiments. \cite{rf:Jalabert,rf:Baranger,rf:BJS,rf:Ketzmerick,rf:Takane1}
An interesting result that has emerged concerns the magnetotransport of doubly connected ballistic 
cavities, i.e., Aharonov-Bohm (AB) cavities \cite{rf:Kawabata1,rf:Kawabata2,rf:Kawabata3,rf:Taylor} (see Fig. 1).
We have calculated the $average$
conductance for these systems and showed that the self-averaging effect causes the $h/2e$ 
Altshuler-Aronov-Spivak (AAS) oscillation, \cite{rf:Kawabata1} which is ascribed
to interference between time-reversed coherent back-scattering classical trajectories.
Moreover we have showed that the AAS oscillation in these systems
becomes an experimental probe of the quantum chaos. 
Another interesting phenomenon in these systems is the $h/e$ AB oscillation 
for $nonaveraged$ conductance.
The result of numerical calculations \cite{rf:Kawabata2} indicated that the period of the energy averaged conductance,
\begin{equation}
     <g(\phi)>_E =  \frac{1}{\Delta E} \int _{E_F -\Delta E /2} ^{E_F + \Delta E /2} g(E,\phi) dE
	 ,
\end{equation}
changed from $h/2e$
to $h/e$, when the range of energy average $\Delta E$ is decreased.
However, little is known about the effect of chaos on the $h/e$ AB oscillation in AB cavities.
In this paper, we shall calculate the correlation function $C(\Delta \phi)$ of the $nonaveraged$ conductance
by using the semiclassical theory
and show that $C(\Delta \phi)$ is qualitatively different between 
chaotic and regular AB cavities.  

In the following, we shall derive $C(\Delta \phi)$ separately for chaotic and regular
AB cavities in which uniform normal magnetic 
field $B$ (AB flux) penetrates only through the hollow. 
%
%
%
%
%
%
\begin{center}
\begin{figure}[b]
\hspace{0.3cm}
\epsfxsize=8.5cm
\epsfbox{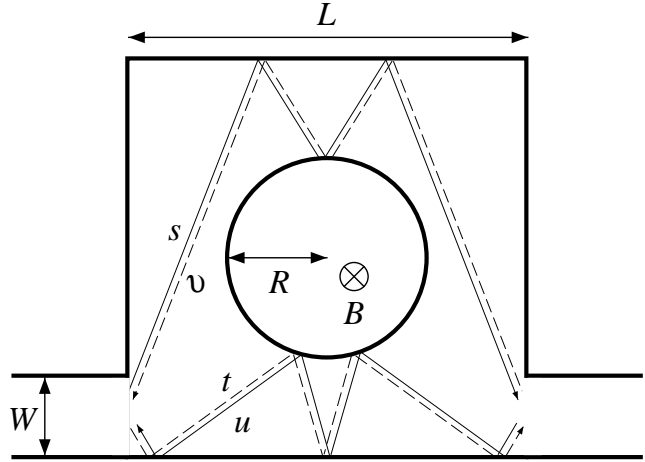}
\caption{An example of the pair of 
four classical paths which contribute to the correlation function for the Aharonov-Bohm cavity.
}
\end{figure}
\end{center}
%
%
%
%
The  transmission amplitude from a mode $m$ on the left to a mode $n$ on the right
for electrons at the Fermi energy is given by \cite{rf:FL}
\begin{equation}
    t_{n,m}  =  - i \hbar \sqrt{\upsilon_n \upsilon_m} 
    \int dy \int dy' \psi_n^*(y') \psi_m(y) 
  G(y',y,E_F)
  ,
  \label{eqn:e4-4-2}
\end{equation}
where \(\upsilon_m(\upsilon_n)\) and \(\psi_m(\psi_n)\) are the longitudinal velocity and transverse 
wave function for the mode $m$ ($n$) at a pair of lead wires attached to the billiards.  
In eq.~(\ref{eqn:e4-4-2}), $G$ is the retarded Green's function.
In order to carry out the semiclassical approximation, we replace $G$ by the semiclassical Green function, \cite{rf:Gutzwiller}
\begin{equation}
  G^{sc}(y',y,E) = \frac {2 \pi} {(2 \pi i \hbar)^{3/2}} \sum_{s(y,y')} 
  \sqrt{D_s} 
  \exp \left[ 
                            \frac i {\hbar} S_s (y',y,E) - i \frac \pi {2} \mu_s
       \right]
  \label{eqn:e4-4-4}
\end{equation}
where $S_s$ is the action integral along a classical path $s$,
the pre-exponential factor is
\begin{equation}
  D_s = \frac{ m_e } {\upsilon_F \cos{\theta'}} 
             \left| 
                \left(
			       \frac{\partial  \theta }{\partial y' }
			    \right)_{y}
			 \right|
  \label{eqn:e4-4-5}
\end{equation}
with $\theta$ and $\theta'$ the incoming and outgoing angles, respectively, and $\mu$ is the Maslov 
index.  
Substitute eq.~(\ref{eqn:e4-4-4}) into eq.~(\ref{eqn:e4-4-2})
and carrying out the double integrals by the saddle-point approximation, we obtain  
\begin{equation}
  t_{n,m} = - \frac {\sqrt{2 \pi i \hbar}} {2 W} \sum_{s(\bar n,\bar m)} 
  {\rm sgn} (\bar n) {\rm sgn} (\bar m) \sqrt{\tilde D_s}
   \exp{ 
        \left[
              \frac i {\hbar} \tilde S_s (\bar n,\bar m;E)-i \frac \pi {2} \tilde \mu_s
        \right]
	  }
	  ,
  \label{eqn:e6-2-2}
\end{equation}
where $W$ is the width of the hard-wall leads and \( \bar m = \pm m \).
In eq.~(\ref{eqn:e6-2-2}), 
\( 
  \tilde{S_s} (\bar{n},\bar{m};E) = S_s(y'_0,y_0;E)+ \hbar \pi ( \bar{m} y_0 - \bar{n} y'_0 ) / W 
\), 
\( 
  \tilde{D_s} = ( m_e \upsilon_F \cos{\theta'})^{-1} \left| ( \partial y /\partial \theta' )_{\theta} \right|
\) 
and
\( 
\tilde{\mu_s} = \mu_s + H \left( -( \partial \theta / \partial y )_y' \right)
                      + H \left( -( \partial \theta' / \partial y' )_{\theta} \right),
\)
respectively, where $\theta=\sin^{-1}(\bar{n} \pi / k W )$ and $H$ is the Heaviside step function.

Transmission coefficients between modes are obtained by taking the absolute square of 
transmission amplitudes, $T_{n,m}=\left| t_{n,m} \right|^2$.
For leads of width $W$ that support $N_M= \mbox{Int} [kW/\pi]$ modes, the total transmitted intensity
summed over $m$ and $n$ is
\begin{equation}
	  T(k)= \frac{1}{2} \frac{\pi}{kW} 
				  \sum_{n,m}^{N_M} \sum_{s,u} 
				        \sqrt{\tilde A_s \tilde A_u}   
	                    \exp \left[
			                        i k \left(
			                        \tilde L_s - \tilde L_u
			                            \right)
			                        +i \pi \nu_{s,u}
			                 \right]
	  ,
  \label{eqn:e6-2-3}
\end{equation}
where $s$ and $u$ label the classical trajectories.
In eq.~(\ref{eqn:e6-2-3}),
\( \tilde L_s = \tilde S_s / k \hbar \)
,
\( \nu_{s,u} = \left( \tilde \mu_u - \tilde \mu_s \right) / 2 \)
, and
\( \tilde A_s = \left( \hbar k / W \right) \tilde D_s \).
The fluctuations of the conductance $g=(e^2/\pi \hbar) T(k)$ are defined by their deviation from the classical
value; in the absence of any symmetries,
%
%
\begin{equation}
 \delta g \equiv g - g_{cl} 
 .
\label{eqn:e6-2-9}
\end{equation}
%
%
In this equation $g_{cl} = (e^2/\pi \hbar) T_{cl}$, where $T_{cl}$ is the classical total transmitted intensity.
In order to characterize the $h/e$ AB oscillation, we define the correlation function of the oscillation in 
magnetic field $B$ by the average over $B$,
%
%
\begin{equation}
 C(\Delta B) \equiv \left< \delta g(B) \delta g(B+\Delta B)\right>_{B} 
 .
\label{eqn:e6-2-10}
\end{equation}
%
%
With use of the ergodic hypothesis, $B$ averaging can be replaced by the $k$ averaging, i.e.,
%
%
\begin{equation}
 C(\Delta B) = \left< \delta g(k,B) \delta g(k,B+\Delta B) \right>_{k} 
 .
\label{eqn:e6-2-11}
\end{equation}
%
%
Within the diagonal approximation \cite{rf:Jalabert,rf:BJS} the correlation function of transmission coefficients between the modes is given by
%
%
\begin{equation}
 C_D (\Delta B) = \left( \frac{e^2}{\pi \hbar} \right)^2
                  \left<  
				        \sum_{n,m=1}^{N_M} \delta T_{n,m} (k,B) \delta T_{n,m} (k,B+\Delta B) 
				  \right>_{k} 
 ,
\label{eqn:e6-2-12}
\end{equation}
%
%
where $\sum \nolimits_{n,m} \delta T_{n,m} = T(k) - T_{cl}$.
The semiclassical expression for the transmission amplitudes, eq.~(\ref{eqn:e6-2-3}), yields
%
%
\begin{eqnarray}
 C_D (\Delta B) &=& \!\left( \frac{e^2}{\pi \hbar} \right)^2
                  \!
                  \frac{1}{4}
				  \int_{0}^{1} d \sin \theta
				  \int_{0}^{1} d \sin \theta'
				  \!
			      \sum_{s(\bar{\theta},\bar{\theta'})} \sum_{u \ne s}
			      \sum_{t(\bar{\theta},\bar{\theta'})} \sum_{v \ne t}
				  \sqrt{\tilde {A_s}\tilde {A_u}}
				  \sqrt{\tilde {A_t}\tilde {A_v}}
				  e^{i \pi (\nu_{s,u}-\nu_{t,v})} \nonumber\\
				  &&\times
                  \left<
				       \exp
					      \left[
						       \frac{i}{\hbar} \left\{
							                         \tilde {S_s} (B) -  \tilde {S_u} (B) +
													 \tilde {S_t} (B+\Delta B) -  \tilde {S_v} (B+\Delta B)
							                   \right\}
						  \right]
				  \right>_{k} 
 ,
\label{eqn:e6-2-14}
\end{eqnarray}
%
%
where $\bar{\theta}= \pm \theta$.
As for AAS oscillation, \cite{rf:Kawabata1} the diagonal approximation yields an expression with $k$ dependence only in the exponent.
With use of $\tilde {S_s}(B) = \hbar k \tilde {L_s} +e \int_{s} {\bf A} \cdot d{\bf r}$, we get
%
%
\begin{eqnarray}                  
       &&
	   \left<
			\exp
			\left[
				  \frac{i}{\hbar} 
				  \left\{
				      \tilde {S_s} (B) -  \tilde {S_u} (B) +
					  \tilde {S_t} (B+\Delta B) -  \tilde {S_v} (B+\Delta B)
				  \right\}
			\right]
        \right>_{k} \nonumber\\
		&=& 
       \left<
			\exp
			\left[
				  \frac{i}{\hbar} 
				  \left\{
				      \tilde {L_s} -  \tilde {L_u} +
					  \tilde {L_t}  -  \tilde {L_v} 
				  \right\}
			\right]
        \right>_{k}
		\exp
		\left[
		    i
		    \frac{e}{\hbar}
		    \left(
		        \int_{s} 
				\!-\!
				\int_{u} {\bf A} \cdot d{\bf r}
				+
				\int_{t} 
				\!-\!
				\int_{v} {\bf A}' \cdot d{\bf r}
		  \right)
		\right]
        .
\label{eqn:e6-2-15}
\end{eqnarray}
%
%
Here $ 2 \pi \int_{s(u)} {\bf A} \cdot d{\bf r} = B \Theta_{s(u)}$ and $ 2 \pi \int_{t(v)} {\bf A}' \cdot d{\bf r} = (B + \Delta B) \Theta_{t(v)}$.
The finite $k$ average implies that only contribution is expected for 
%
%
\begin{equation}
		\tilde {L_s} -  \tilde {L_u}  +
		\tilde {L_t}  -  \tilde {L_v} 
		=
		0
\label{eqn:e6-2-16}
\end{equation}
%
%
exactly.
Because of the definition of $C_D$ in eq.~(\ref{eqn:e6-2-11}), all four paths satisfy the same boundary conditions for 
angles, and hence they are all chosen from the same discrete set of paths.
In the absence of symmetry, the only contribution is $v=s$ and $t=u$.
The terms with $s=u$ and $t=v$ are excluded because they represent the average values that must be removed from the
correlation functions.
In Fig. 1 we show the typical set of trajectories that contribute to the correlation function.
This process is analogous to the two diffuson propagators in a diffusive regime \cite{rf:UCF}. 
Since magnetic flux penetrates only through the hollow, the exponent in eq.~(\ref{eqn:e6-2-15}) becomes
%
%
\begin{equation}
	i
	\frac{e}{h}
	\Delta B
    (\Theta_u - \Theta_s)
	=
	\pm
	i 
	\frac{\Delta \phi}{\phi_0}
	\left\{
	2 \pi + (w_u -w_s) 
	\right\}
 ,
 \label{eqn:e6-2-18}
\end{equation}
%
%
where $\pm$ corresponds to the clockwise (counterclockwise) rotation to the center disk for path $u$.
In eq.~(\ref{eqn:e6-2-18}) $w_s$ is the winding number of classical path $s$.
Therefore we obtain 
%
%
\begin{eqnarray}
 C_D (\Delta \phi) = \left( \frac{e^2}{\pi \hbar} \right)^2
				  e^{2 \pi i \frac{\Delta \phi}{\phi_0}}
				  \;
				  \left|
                  \frac{1}{2}
				  \int_{0}^{1} d \sin \theta
				  \int_{0}^{1} d \sin \theta'
			      \sum_{s(\bar{\theta},\bar{\theta'})}
				  \tilde {A_s}
				  e^{-2 \pi i w_s \frac{\Delta \phi}{\phi_0}}
				  \right|^2
				  +
				  c.c.
				  \enspace
				  .
\label{eqn:e6-2-20}
\end{eqnarray}
%
%
In order to evaluate sum over $s$ and integrations on $\theta (\theta')$, we shall reorder the trajectories according to the 
increasing dwelling time $T_{s}$.
Therefore we find for the diagonal part of the semiclassical correlation function for chaotic systems as 
%
%
\begin{eqnarray}
 C_D (\Delta \phi) = 
                  C_D(0)
				  \cos \left( 2 \pi \frac{\Delta \phi}{\phi_0} \right)
				  \left\{
                  \frac{\cosh \delta - 1}
                  {\cosh \delta - \cos \left(  2 \pi \frac{\Delta \phi}{\phi_0}\right) }				
                  \right\}^2
				  ,
\label{eqn:e6-2-22}
\end{eqnarray}
%
%
where \( \delta = \sqrt { 2 T_0 \gamma / \alpha } \). 
In deriving eq.~(\ref{eqn:e6-2-22}) we have used the exponential dwelling time 
distribution, $N(T) \sim \exp(- \gamma T)$, \cite{rf:BJS,rf:BS} and the Gaussian winding number distribution for 
fixed $T$,\cite{rf:Berry} i.e.,
%
%
\begin{equation}
P(w;T)  = \sqrt{ \frac{T_0} {2 \pi \alpha T}} 
     \exp \left( -\frac{w^2 T_0 } { 2 \alpha T} \right)
  \label{eqn:b4}
  ,
\end{equation}
%
%
where $T_0$ and $\alpha$ are the system-dependent constants corresponding to
the dwelling time for the shortest classical winding trajectory and 
the variance of the distribution of $w$, respectively.
By using the extended semiclassical theory, \cite{rf:Takane2} we can take account of the 
off-diagonal part and the influence of the small-angle diffraction as
%
%
\begin{eqnarray}
 C (\Delta \phi) =  \left( \frac{e^2}{\pi \hbar} \right)^2 \frac{1}{8} \frac{C_D (\Delta \phi)}{C_D (0)}
				  ,
\label{eqn:e6-2-23}
\end{eqnarray}
%
%
for the case which the widths of the lead wires are equal.
Then we obtain the full correlation function for chaotic AB cavities,
%
%
\begin{eqnarray}
\!\!\! C (\Delta \phi) \!\!&=& \!\!\!\!  \left( \frac{e^2}{\pi \hbar} \right)^2
                  \frac{1}{8}
				  \cos \left( 2 \pi \frac{\Delta \phi}{\phi_0} \right)
				  \left\{
                  \frac{\cosh \delta - 1}
                  {\cosh \delta - \cos \left(  2 \pi \frac{\Delta \phi}{\phi_0}\right) }
                  \right\}^2 \nonumber\\
				\!\! &=& \!\!\!\!
				  \left( \frac{e^2}{\pi \hbar} \right)^2
                  \frac{1}{8}
				  \left(
				  \frac{\cosh \delta \!-\! 1}{\sinh \delta}
				  \right)^2
				  \!
				  \cos \left( 2 \pi \frac{\Delta \phi}{\phi_0} \right)
                  \left\{
                     1 \!+\! 2 \sum_{n=1}^{\infty} e^{-\delta n}
			         \cos \left( 2 \pi n \frac{\Delta \phi} {\phi_0} \right)
                  \right\}^2 
				  .
\label{eqn:e6-2-24}
\end{eqnarray}
%
%
The periodic function $C (\Delta \phi)$ has the minimum value 
%
%
\begin{eqnarray}
 \frac{C (\Delta \phi_{min})}{C (0)}
       =\left(
	        \frac{\cosh \delta - 1}{\cosh \delta + 1} 
	    \right)^2
\label{eqn:e6-2-25}
\end{eqnarray}
%
%
at $\Delta \phi_{min} = n \pi$, where $n=1,3,5,\cdots$.
Therefore, $C (\Delta \phi)$ oscillates with the period $\phi_0$, i.e., $AB$ $oscillation$.
From the above results, we can conclude that it is possible to predict quantitatively $C (\Delta \phi)$ of the chaotic AB cavities
from a knowledge of the chaotic classical scatterings dynamics.
Note for consistency that the field scale of fluctuations is twice that of AAS oscillation \cite{rf:Kawabata1}
because the relevant phase involves the difference between two winding numbers whereas
AAS oscillation involves the sum. 
Surprisingly the semiclassical formula, eq.~(\ref{eqn:e6-2-24}), is quite similar to Isawa $et$ $al.$'s results 
for the $disordered$ quasi-one dimensional AB ring: \cite{rf:Isawa} 
%
%
\begin{equation}
  C(\Delta \phi) = \frac{e^4}{\hbar^2}
                16
                \frac{L_{\varphi}} {2 \pi R}
                \frac{2 + \cos \left( 2 \pi \frac{\Delta \phi}{\phi_0} \right)} 
                { 
			     \left[   
			          \cosh \left(
						         \frac{2 \pi R}{L_{\varphi}}  
							\right) 
					  - 
					  \cos \left( 2 \pi \frac{\Delta \phi}{\phi_0} \right) + Q 
			     \right] ^2
			    }
				.
\label{eqn:e6-2-26}
\end{equation}
%
%
In this equation $L_{\varphi}$ is the phase coherence length, $R$ is the radius of the ring and 
$Q=\sinh(2 \pi R / L_{\varphi}) + \sinh^2(\pi R / L_{\varphi}) /2$, respectively.

The periodic function $C(\Delta \phi)$ is large and positive for very small $\Delta \phi$,
and has the limiting value
%
%
\begin{eqnarray}
 C (0) = \left( \frac{e^2}{\pi \hbar} \right)^2
                  \frac{1}{8}
				  .
\label{eqn:e6-2-27}
\end{eqnarray}
%
%
This result is consistent with the result of random matrix theory for the circular orthogonal ensemble. \cite{rf:BM,rf:JPB}
In the case of weak $\Delta \phi$, eq.~(\ref{eqn:e6-2-24}) is rewritten asymptotically as
%
%
\begin{eqnarray}
 \frac{C (\Delta \phi)}{C (0)}
       \approx
	   1-
	   2 \pi^2
       \left(
	        \frac{\cosh \delta + 1}{\cosh \delta - 1} 
	   \right)
	   \Delta \phi^2
		.
\label{eqn:e6-2-28}
\end{eqnarray}
%
%
Therefore $C (\Delta \phi)$ decreases quadratically with increasing $\Delta \phi$ near 
$\Delta \phi =0$.
The quadratic behavior of $C (\Delta \phi)$ is similar to that for ordinal chaotic cavity, $e.g.$, stadium,
at near $\Delta B =0$. \cite{rf:Jalabert,rf:BJS}

On the other hand, for the regular cases, we use $N(T) \sim T^{-\beta}$ \cite{rf:BJS,rf:Lai} in eq.~(\ref{eqn:e6-2-20})
Assuming as well the Gaussian distribution of $P(w;T)$, we get
%
%
\begin{eqnarray}
 C (\Delta \phi) = C (0)
				  \cos \left( 2 \pi \frac{\Delta \phi}{\phi_0} \right)
				  \left\{
                     \frac{
                     1+2 \displaystyle{\sum_{n=1}^{\infty}
                      F  \left( \beta-\frac1{2}, \beta+\frac1{2};-\frac{n^2}{2\alpha}  \right)
                      \cos \left( 2 \pi n \frac{\Delta \phi} {\phi_0} \right) }
					  }
					  {
                     1+2  \displaystyle{\sum_{n=1}^{\infty} 
                     F  \left( \beta-\frac1{2}, \beta+\frac1{2};-\frac{n^2}{2\alpha}  \right)}
 					  }                  
				  \right\}^2
				  ,
\label{eqn:e6-2-29}
\end{eqnarray}
%
%
where $F$ is the hyper-geometric function of confluent type.
As in AAS oscillation, parameters $\beta$ and $\alpha$ characterizing the classical dynamics
determine the behavior of $C(\Delta \phi)$.

 Next we shall see the difference of $C(\Delta \phi)$ for chaotic and regular AB cavities in detail.
In the chaotic AB cavity, a main contribution to the AB oscillation comes from the $n=1$ component.
Figure 2 shows (a) aspect ratio ($\sigma=R/W$) and (b) the degree of opening to the lead 
wires ($\eta=L/W$) dependence of $C(\Delta \phi)$ for the open chaotic AB cavity (Sinai billiard \cite{rf:Sinai}),
where $R$ is the radius of the center circle and $L$ is the linear dimension of the outer square. 
The classical parameter $\delta$ is calculated by using geometric dimensions of the cavity. \cite{rf:Kawabata1,rf:Kawabata2}
In the case of small $\sigma$ or $\eta$, classical trajectories are able to wind around the center disk many times and 
the higher harmonics can contribute to AB oscillation.
Therefore one can see from Fig. 2 that the minimum value $C(\Delta \phi_{min})/C(0)$ slightly increase from -1
as $\sigma$ or $\mu$ becomes small.

%
%
%
%
%
\begin{center}
\begin{figure}[t]
\hspace{3.3cm}
\epsfxsize=8.5cm
\epsfbox{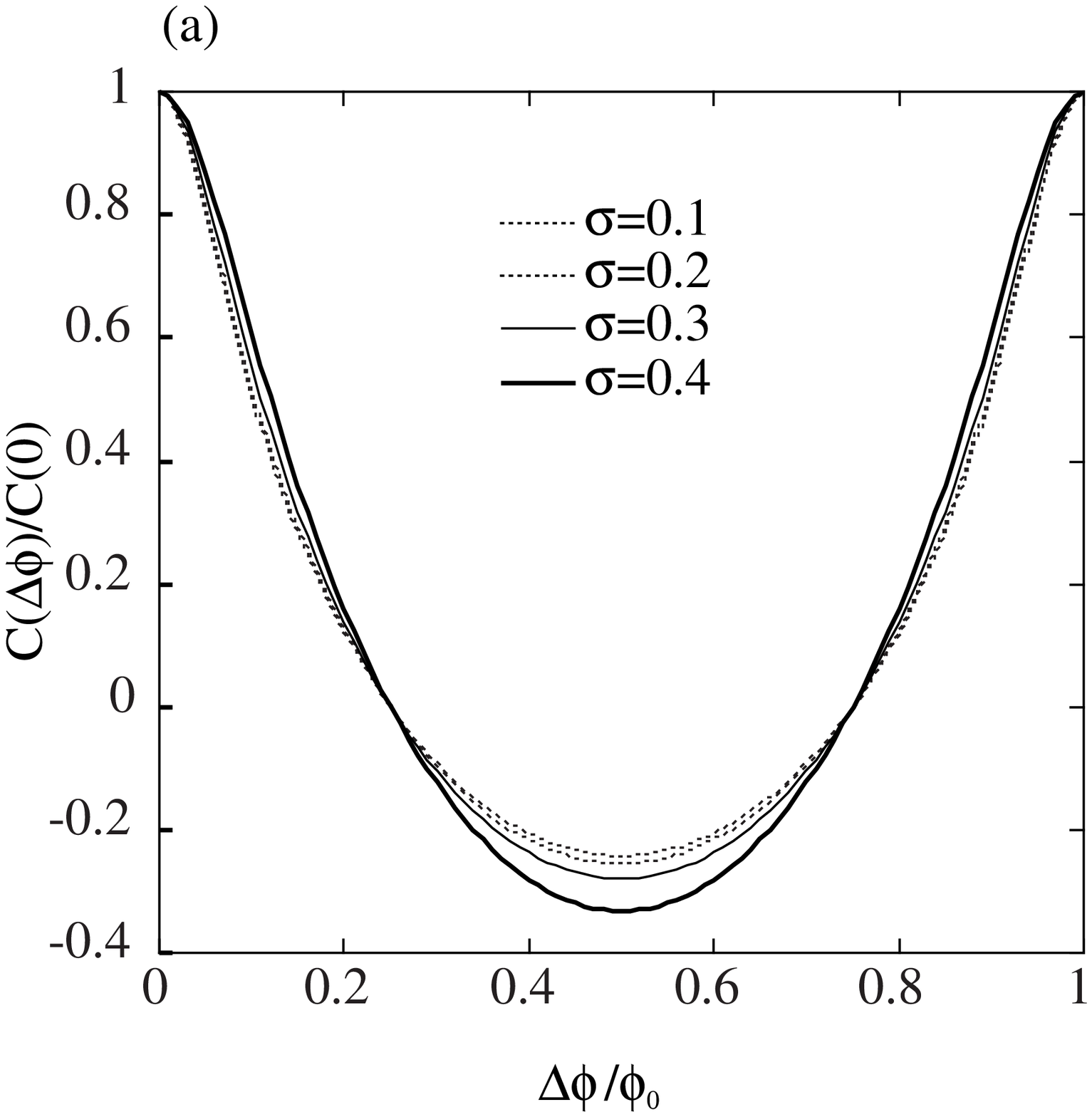}
\epsfxsize=8.5cm
\epsfbox{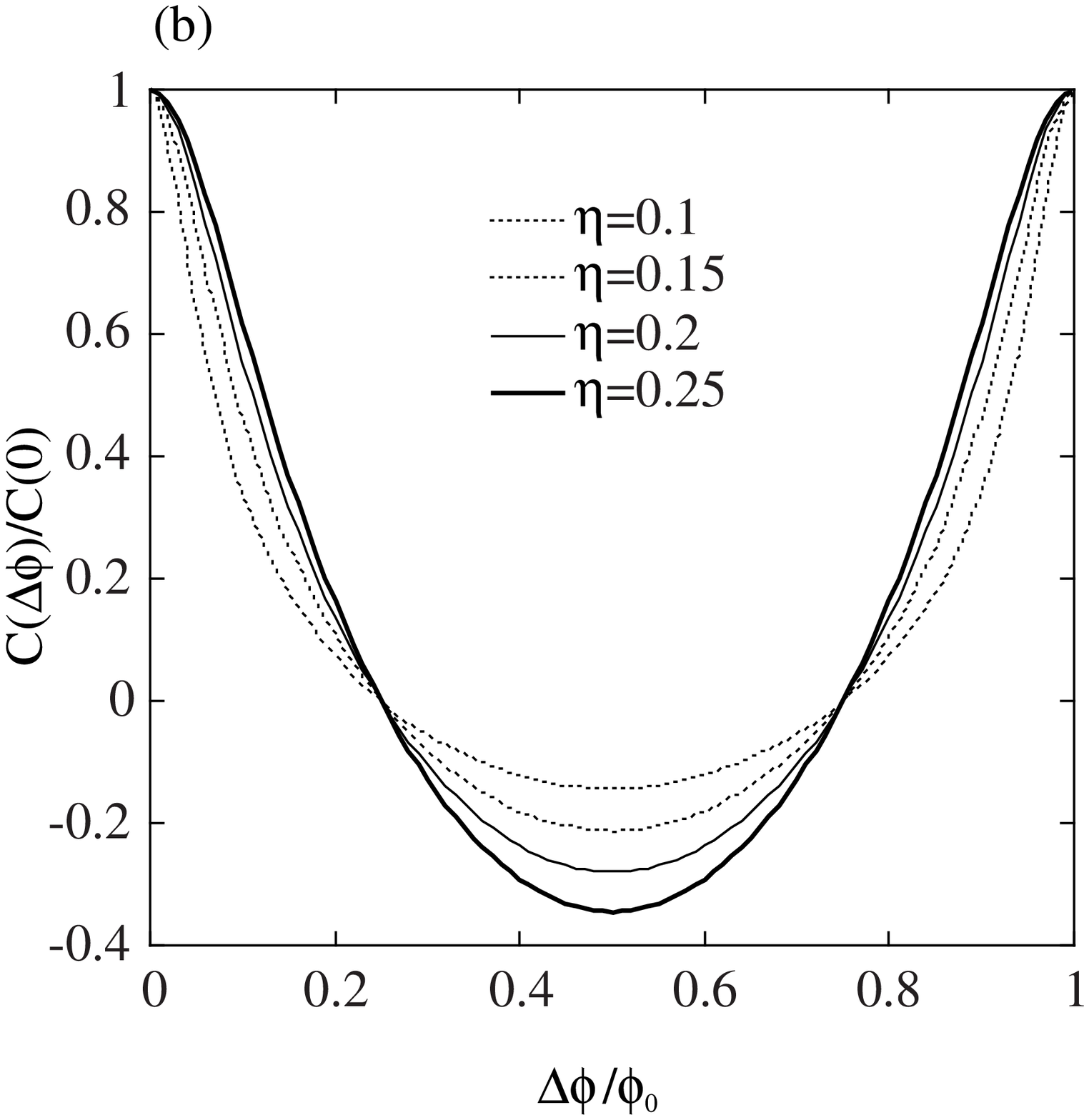}
\caption{
Semiclassical correlation function $C(\Delta \phi)$ of the conductance as a function of $\Delta \phi$
for the chaotic AB (Sinai) cavity:
(a) for various aspect ratio $\sigma=R/L$ ($\eta=0.2$)
;
(b) for various degrees of opening $\eta=W/L$ ($\sigma=0.3$).
}
\end{figure}
\end{center}
%
%
%
%

On the other hand, for regular cases, the amplitude of the AB oscillation decays algebraically, i.e., $F \sim n^{-2 \beta -1}$
for large $n$. 
This behavior is caused by the power law dwelling time distribution, i.e., $N(T) \sim T^{-\beta}$. 
Thus, in contrast to the chaotic cases, we can expect that the considerably higher harmonics contribution
causes a noticeable deviation from the cosine function for $C(\Delta \phi)$.
Therefore, between the difference $C(\Delta \phi)$ of these ballistic AB cavities
can be attributed to the difference of chaotic
and regular classical scattering dynamics.

In summary, we have investigated magnetotransport in single ballistic cavities whose structures form AB geometry by use 
of semiclassical methods with a particular emphasis on the derivation of the 
semiclassical formulas. 
The existence of the AB oscillation of $nonaveraged$ magnetoconductance is predicted for single chaotic and 
regular AB cavities. 
Furthermore, we find that the difference between classical dynamics leads to qualitatively different behaviors
for the correlation function. 
The AB oscillation in the ballistic regime will provide a new experimental testing ground for exploring quantum chaos.

We would like to acknowledge 
K. Nakamura and Y. Takane
for valuable discussions and comments.


\begin{references}
%
%
%
%
\bibitem{rf:QC}
For a review of quantum chaos in mesoscopic systems see 
{\it Chaos and Quantum Transport in Mesoscopic Cosmos} 
, edited by K. Nakamura,
Chaos Solitons Fractals 
{\bf 8}, No.7/8 (1997).
%
%
%
%
\bibitem{rf:Jalabert}
R.A. Jalabert, H.U. Baranger and A.D. Stone,
Phys. Rev. Lett. {\bf 65}, (1990) 2442.
%
%
%
%
\bibitem{rf:Baranger}
H.U. Baranger, R.A. Jalabert and A.D. Stone,
Phys. Rev. Lett. {\bf 70}, (1993) 3876.
%
%
%
%
\bibitem{rf:BJS}
H.U. Baranger, R.A. Jalabert and A.D. Stone,
Chaos {\bf 3}, (1993) 665.
%
%
%
%
\bibitem{rf:Ketzmerick}
R. Ketzmerick,
Phys. Rev. B {\bf 54}, (1996) 10841.
%
%
%
%
\bibitem{rf:Takane1}
Y. Takane and K. Nakamura,
J. Phys. Soc. Jpn. {\bf 66}, (1997) 2977.
%
%
%
%
\bibitem{rf:Kawabata1}
S. Kawabata and K. Nakamura,
J. Phys. Soc. Jpn. {\bf 65}, (1996) 3708.
%
%
%
%
\bibitem{rf:Kawabata2}
S. Kawabata and K. Nakamura,
Chaos Solitons Fractals {\bf 8}, (1997) 1085. 
%
%
%
%
\bibitem{rf:Kawabata3}
S. Kawabata and K. Nakamura,
Phys. Rev. B {\bf 57}, (1998) 6282.
%
%
%
%
\bibitem{rf:Taylor}
R.P. Taylor, R. Newbury, A.S. Sachrajda, Y. Feng, P.T. Coleridge, C. Dettmann,
N. Zhu, H. Guo, A. Delage, P.J. Kelly, and Z. Wasilewski,
Phys. Rev. Lett. {\bf 78}, (1997) 1952;
R.P. Taylor, A.P. Micolich, R. Newbury, and T.M. Fromhold,
Phys. Rev. B {\bf 56}, (1997) R12733;
A.S. Sachrajda, R. Ketzmerick, C. Gould, Y. Feng, P.J. Kelly, A. Delage, and Z. Wasilewski,
Phys. Rev. Lett. {\bf 80}, (1998) 1948.
%
%
%
%
\bibitem{rf:FL}
D.S. Fisher and P.A. Lee, 
Phys. Rev. B {\bf 23}, (1981) 6851.
%
%
%
%
\bibitem{rf:Gutzwiller}
M.C. Gutzwiller, {\it Chaos in Classical and Quantum Mechanics}, 
(Springer, New York, 1990).
%
%
%
%
\bibitem{rf:UCF}
B.L. Altshuler and D.E. Khmelnitskii,
JETP Lett. {\bf 42}, (1985) 359; 
P.A. Lee, A.D. Stone, and H. Fukuyama, 
Phys. Rev. B {\bf 35}, (1987) 1039.
%
%
%
%
\bibitem{rf:BS}
R. Bl\"umel and U. Smilansky,
Phys. Rev. Lett. {\bf 60}, (1988) 477.
%
%
%
%
\bibitem{rf:Berry}
M.V. Berry and J.P. Keating,
J. Phys. A {\bf 27}, (1994) 6167.
%
%
%
%
\bibitem{rf:Takane2}
Y. Takane and K. Nakamura,
J. Phys. Soc. Jpn. {\bf 67}, (1998) 397.
%
%
%
%
\bibitem{rf:Isawa}
Y. Isawa, H. Ebisawa and S. Maekawa, 
J. Phys. Soc. Jpn. {\bf 55}, (1986) 2523;
in {\it Proc. 2nd Int. Symp. Foundations of Quantum Mechanics},
eds. M. Namiki, Y. Ohnuki
, Y. Murayama and S. Nomura (Physical Society of Japan, Tokyo, 1987) pp.218.
%
%
%
%
\bibitem{rf:BM}
H.U. Baranger and P.A. Mello,
Phys. Rev. B {\bf 51}, (1995) 4703.
%
%
%
%
\bibitem{rf:JPB}
R.A. Jalabert, J.-L. Pichard and C.W.J. Beenakker,
Europhys. Lett. {\bf 27}, (1994) 255.
%
%
%
%
\bibitem{rf:Lai}
Y.C. Lai, R. Bl\"umel, E. Ott and C. Grebogi, 
Phys. Rev. Lett. {\bf 68}, (1992) 3491.
%
%
%
%
\bibitem{rf:Sinai}
Y.G. Sinai,
Russ. Math. Surv. {\bf 25}, (1979) 137. 
%
%
%
%
\end{references}
\end{document}